\documentclass[aps,pre,onecolumn,superscriptaddress,showkeys]{revtex4}
\usepackage[english]{babel}
\usepackage{amsmath}
\usepackage{graphicx}
\usepackage[retainorgcmds]{IEEEtrantools}
\newcommand{\rd}{\mathrm{d}}
\newcommand{\nl}{\nonumber \\}
\newcommand{\mc}{\IEEEeqnarraymulticol}
\newcommand{\la}{\langle}
\newcommand{\ra}{\rangle}
\newcommand{\eps}{\varepsilon}

\newcommand{\dlta}{\mathrm{\Delta}}

\begin{document}

\title{Geometrical detection of weak non-Gaussianity upon coarse-graining}


\author{T.H. Beuman}
  \affiliation{Instituut-Lorentz for Theoretical Physics, Leiden University, NL 2333 CA Leiden, The Netherlands}
\author{Ari M. Turner}
  \affiliation{Department of Physics and Astronomy, Johns Hopkins University, Baltimore, MD 21218, USA}
\author{V. Vitelli}
  \email{vitelli@lorentz.leidenuniv.nl}
  \affiliation{Instituut-Lorentz for Theoretical Physics, Leiden University, NL 2333 CA Leiden, The Netherlands}

\begin{abstract}

Measures of the non-Gaussianity of a random field depend on how accurately one is able to measure the field. If a signal measured at a certain point is to be averaged with its surroundings, or coarse-grained, the magnitude of its non-Gaussian component can vary. In this article, we investigate the variation of the ``apparent'' non-Gaussianity, as a function of the coarse-graining length, when we measure non-Gaussianity using the statistics of extrema in the field. We derive how the relative difference between maxima and minima -- which is a geometrical measure of the field's non-Gaussianity -- behaves as the field is coarse-grained over increasingly larger length scales. Measuring this function can give extra information about the non-Gaussian statistics and facilitate its detection.


\end{abstract}

\maketitle

\section{Introduction}
\label{sec:intro}

As a consequence of the central limit theorem or of linear physics, many random fields are -- at least as a first-order approximation -- almost Gaussian. Examples of such fields can be found in various disciplines. The cosmic microwave background \cite{cite:Dodelson} is widely studied, but examples can also be found in optical speckle fields \cite{cite:Flossmann} and maps of brain activity \cite{cite:Worsley}. Gaussian fields have universal properties that are well studied, including the statistics of its extrema and other singular points such as umbilics \cite{cite:Longuet1,cite:Longuet2,cite:Berry,cite:Dennis1,cite:Dennis2}.

Deviations from Gaussianity are indicative of underlying nonlinear processes. Distilling these non-Gaussianities from the main signal can thus shed light on these nonlinear mechanisms. A standard way of testing and quantifying non-Gaussianity is to determine correlation functions beyond the second order. For Gaussian fields, these should, by Wick's theorem, be factorizable into two-point correlation functions. A mismatch in these relations is therefore a tell-tale sign of non-Gaussianity.

However, this method requires detailed measurements of the field in question, and may therefore be impractical in some experimental settings. In \cite{cite:paper_local,cite:paper_kpz,cite:paper_PNAS}, an alternative method was introduced, based on looking at the statistics of extrema and umbilical points, which in some circumstances can be less dependent on the accuracy with which the field can be probed. Removing one of these geometrical singularities or topological defects typically involves a nonlocal operation on the field that is unlikely to arise from local sources of noise or other perturbations.

In the present work, we investigate the effects that coarse-graining has on the statistics of extrema of a non-Gaussian field. This coarse-graining can represent the effect of imprecise measurements or averaging, but it can sometimes be helpful for the geometrical detection of non-Gaussianities. As illustrated in \cite{cite:paper_local}, a local non-Gaussian perturbation actually has an insignificant effect on the statistics of maxima and minima of the field if the resolution of the measurement is perfect. As a result, the perturbation would thus go undetected. As will be shown here, coarse-graining such a perturbed field can give rise to a sizable imbalance, thereby bringing the non-Gaussianity to light. For nonlocal perturbations, there is an imbalance at any resolution, but by comparing the number of maxima and minima at different scales one could get more information about the nonlocal correlations.

An example in which coarse-graining plays a different but intriguing role is gravitational lensing tomography. Images from distant galaxies are sheared due to mass present between that galaxy and us -- this phenomenon is called weak gravitational lensing. Measuring the shear offers a window on the distribution of mass in the universe \cite{cite:Hoekstra} -- the singularities of the shear field for instance correspond to the umbilics of the projected gravitational potential \cite{cite:Vitelli}. It is a two-dimensional window though, providing only information about the projected gravitational potential. If one however incorporates the redshift -- which can be translated to distance -- a three-dimensional picture can be constructed \cite{cite:Hu}. The entire range of redshifts is divided into bins that can be processed individually. Therefore each bin yields an average of sources from a range of distances, rather than one specific distance. This can be interpreted as an effective source of coarse-graining.

The outline of this paper is as follows. The basics of Gaussian and non-Gaussian fields are given in section~\ref{sec:gsnfields}. Section~\ref{sec:coarsegrain} sets up the mathematical framework for coarse-graining and discusses its effects. In section~\ref{sec:limit} we derive what the imbalance between maxima and minima looks like in the limit that one coarse-grains over a large scale. This is demonstrated in section~\ref{sec:example}, using an example case that allows the imbalance to be determined for arbitrary coarse-grain length scales. In section~\ref{sec:num}, the analytical result is compared to numerical simulations. Finally, section~\ref{sec:concl} summarizes our findings.

\section{Gaussian and non-Gaussian fields}
\label{sec:gsnfields}

\subsection{Gaussian fields}
\label{subsec:gsnfields}

A homogeneous and isotropic Gaussian field is defined in terms of its Fourier components as
\begin{equation}
  H(\vec{r})  =  \sum_{\vec{k}} A(k) \cos(\vec{k} \cdot \vec{r} + \phi_{\vec{k}}).
  \label{eq:gaussian}
\end{equation}
The phases $\phi_{\vec{k}}$ are independent random variables, uniformly distributed between $0$ and $2\pi$. The \emph{amplitude spectrum} $A(k)$ depends only on the magnitude of the wave vector $\vec{k}$, not its angle -- this ensures isotropy. There are no further constraints on the spectrum. Fields with different spectra can have different characteristics, but all still classify as Gaussian, which gives them some universal properties.
For convenience, we will consider $H$ to be normalized, such that
\begin{equation}
  \la H^2 \ra = \la H(\vec{r})^2 \ra = \bigg( \prod_{\vec{k}} \int \! \frac{\rd \phi_{\vec{k}}}{2\pi} \bigg) \, H(\vec{r})^2 = 1.
\end{equation}

A derivative of a Gaussian field is itself a Gaussian field, as the following example demonstrates:
\begin{equation}
  \frac{\partial}{\partial x} H(\vec{r}) = \sum_{\vec{k}} A(k) k_x \cos(\vec{k} \cdot \vec{r} + \phi_{\vec{k}} - \frac{\pi}2).
\end{equation}
The factor $k_x$ can be absorbed into the spectrum and the $\pi/2$ into the phase, so that the result still conforms to Eq.~\eqref{eq:gaussian}, although the field would no longer be isotropic.

Correlations between Gaussian fields can be completely characterized by the two-point correlations -- higher-order correlations can be factorized into second-order ones as per Wick's theorem.

Equivalent to correlations are the so-called cumulants. The third-order cumulant of three variables for example can be expressed as
\begin{equation}
  \begin{split}
    C_3(X_1, X_2, X_3) =\:	& \la X_1 X_2 X_3 \ra - \la X_1 \ra \la X_2 X_3 \ra - \la X_2 \ra \la X_3 X_1 \ra - \la X_3 \ra \la X_1 X_2 \ra \\
    				& + 2 \la X_1 \ra \la X_2 \ra \la X_3 \ra.
  \end{split}
\end{equation}
Gaussian variables have the characteristic property that all cumulants beyond the second order are zero.

\subsection{Non-Gaussian fields}
\label{subsec:nongsn}

In what follows, an uppercase $H$ is used to designate a Gaussian field and a lowercase $h$ for a perturbed Gaussian field. However, $h$ is always taken to still be homogeneous and isotropic.

There are various types of perturbation. One example is $h(\vec{r}) = H(\vec{r}) + f(H(\vec{r}))$, where $f$ is a (small-valued) function that depends on the original value of $H$ at $\vec{r}$ only. For this reason, this is called a \emph{local} perturbation \cite{cite:paper_local}. When the perturbation depends for instance on the gradient of $H$, the value of $h(\vec{r})$ also encodes information about the surroundings of $\vec{r}$, which is why this type of perturbation is called \emph{nonlocal}.

\section{Coarse-graining}
\label{sec:coarsegrain}

In general, mathematically, coarse-graining a field $h(\vec{r})$ can be expressed as
\begin{equation}
  \tilde{h}(\vec{r}) = \int \rd^2\vec{u} \, K(\vec{u}) h(\vec{r}+\vec{u}),
\end{equation}
with $\int \rd^2\vec{u} \, K(\vec{u}) = 1$.

When applied to a Gaussian field $H(\vec{r})$, the result is (in complex notation):
\begin{equation}
  \begin{split}
    \tilde{H}(\vec{r})	& = \int \rd^2\vec{u} \, K(\vec{u}) \sum_{\vec{k}} A(k) e^{i(\vec{k} \cdot (\vec{r}+\vec{u}) + \phi_{\vec{k}})} \\
    			& = \sum_{\vec{k}} A(k) \Big( \int \rd^2\vec{u} \, K(\vec{u}) e^{i \vec{k} \cdot \vec{u}} \Big) e^{i(\vec{k} \cdot \vec{r} + \phi_{\vec{k}})}
  \end{split}
  \label{eq:gsncg}
\end{equation}
It is thus easily seen that coarse-graining only affects the amplitude spectrum of $H$, but not its Gaussianity.

Consider now what happens when applied to a non-Gaussian field $h$. Let us set $K(\vec{r}) = f(r/l)/l^2$, where $l$ is a length scale that controls the size of the coarse-graining, while $f(x)$ is a dimensionless function that goes to zero for $x \gg 1$. Let $\xi$ be (a measure of) the correlation length of $h$, so that $\la h(\vec{r}) h(\vec{r}+\vec{R}) \ra$ vanishes when $|\vec{R}| \gg \xi$. If $l \gg \xi$, the coarse-graining effectively entails taking the average over a large number (of the order of $(l/\xi)^2$) of independent regions, causing $\tilde{h}$ to acquire Gaussian characteristics on account of the central limit theorem. 

There is a link between coarse-graining and the deterministic Kardar-Parisi-Zhang (KPZ) equation that was investigated in \cite{cite:paper_kpz}. This equation is a diffusion equation that reads
\begin{equation}
  \frac{\partial h}{\partial t} = \nu \nabla^2 h + \frac{\lambda}2 (\nabla h)^2.
\end{equation}
Following the substitution $u = \exp((\lambda/2\nu) h)$, this transforms into $\frac{\partial u}{\partial t} = \nu \nabla^2 u$, with the solution
\begin{equation}
  u(\vec{r},t) = \int \rd^2\tilde{\vec{r}} \, \frac1{4 \pi \nu t} e^{ -\frac{ (\vec{r}-\tilde{\vec{r}})^2 }{ 4 \nu t } } u(\tilde{\vec{r}},0).
\end{equation}
This relation has the same structure as that of the formula for coarse-graining: $u(\vec{r},0)$ can be identified as the original field and $l = \sqrt{\nu t}$ as the coarse-graining scale. The dimensionless coarse-graining function is therefore
\begin{equation}
  f(\rho) = \frac1{4\pi} e^{-\rho^2/4}.
  \label{eq:kpzcg}
\end{equation}

In summary, coarse-graining a non-Gaussian field over a large scale causes it to obtain Gaussian characteristics. A Gaussian field remains Gaussian, regardless of the scale over which it is coarse-grained.

\section{Large scale limit}
\label{sec:limit}


In this section, the consequences of coarse-graining a non-Gaussian field $h$ are investigated, in the limit that the coarse-graining scale becomes very large, as compared to the typical correlation length of the field. In particular, focus is put on the consequences for the densities of maxima and minima.

For a Gaussian field, the densities of maxima and minima are the same due to symmetry. For a non-Gaussian field, this is in general not the case. In \cite{cite:paper_kpz,cite:paper_PNAS}, a general expression was derived for the relative difference between the two densities for a perturbed Gaussian field, up to first order in the perturbation,

\begin{equation}
  \dlta n \equiv \frac{n_{max}-n_{min}}{n_{max}+n_{min}} = \sqrt{\frac6{\pi\alpha}} \Big( \frac43\frac{\beta}{\sigma} + \frac49\frac{\gamma}{\alpha} - \frac{10}{27}\frac{\delta}{\alpha} \Big).
  \label{eq:maxmin}
\end{equation}

The parameters are second- and third-order correlations,
\begin{subequations}
  \begin{align}
    \sigma	& = \la h_z h_{z^*} \ra = C_2(h_z, h_{z^*}),  \label{eq:sigma_def} \\
    \alpha	& = \la h_{zz^*}^2 \ra = C_2(h_{zz^*}, h_{zz^*}),  \label{eq:alpha_def} \\
    \beta	& = \la h_z h_{z^*} h_{zz^*} \ra = C_3(h_z, h_{z^*}, h_{zz^*}),  \label{eq:beta_def} \\
    \gamma	& = \la h_{zz^*}^3 \ra = C_3(h_{zz^*}, h_{zz^*}, h_{zz^*}),  \label{eq:gamma_def} \\
    \delta	& = \la h_{zz} h_{z^*z^*} h_{zz^*} \ra = C_3(h_{zz}, h_{z^*z^*}, h_{zz^*}) \label{eq:delta_def}.
  \end{align}
  \label{eq:correlations}
\end{subequations}
Here the subscripts denote partial differentiation, with $\partial_z = \frac12 (\partial_x - i \partial_y)$ and $\partial_{z^*} = \frac12 (\partial_x + i \partial_y)$. In each cumulant, the variables are all taken at the same point $\vec{r}$, e.g.\ $\sigma = C_2(h_z(\vec{r}) h_{z^*}(\vec{r}))$. Due to homogeneity, the choice of $\vec{r}$ is irrelevant.

Note that it is not generally true that correlations and cumulants are identical. It is true though in this special case of correlations / cumulants up to third order between derivatives of a homogeneous field, as shall be demonstrated with $\gamma$ as an example. Expressing the cumulant in correlations gives
\begin{equation}
  \gamma = C_3(h_{zz^*}, h_{zz^*}, h_{zz^*}) = \la h_{zz^*}^3 \ra - 3 \la h_{zz^*} \ra \la h_{zz^*}^2 \ra + 2 \la h_{zz^*} \ra ^3.
\end{equation}
Note that, with the exception of the first, all terms carry a factor $\la H_{zz^*} \ra$, which can be expressed as
\begin{equation}
  \la h_{zz^*} \ra = \partial_{z_1} \partial_{z_1^*} \la h(\vec{r}) \ra.
\end{equation}
Since we consider $h$ to be homogeneous, $h(\vec{r})$ is constant and thus $\la h_{zz^*} \ra = 0$. This trick applies not only to $\gamma$, but to all five correlations in Eq.~\eqref{eq:correlations}.

For the coarse-grained field $\tilde{h}$, the cumulants can be calculated in the following way (using $\beta$ as an example):
\begin{equation}
  \beta = C_3(\tilde{h}_z, \tilde{h}_{z^*}, \tilde{h}_{zz^*}) = \partial_{z_1}\partial_{z_2^*}\partial_{z_3}\partial_{z_3^*} C_3(\tilde{h}(\vec{r_1}) \tilde{h}(\vec{r_2}) \tilde{h}(\vec{r_3})) \Big|_{\vec{r_1}=\vec{r_2}=\vec{r_3}}
  \label{eq:cg_beta}
\end{equation}
The main ingredients that allow the cumulants to be calculated are thus $C_2(\tilde{h}(\vec{r_1}), \tilde{h}(\vec{r_2}))$ (for $\sigma$ and $\alpha$) and the third-order equivalent (for $\beta$, $\gamma$ and $\delta$).

Let $\xi$ be a measure of the correlation length of $h$, in the sense that $h(\vec{r_1})$ and $h(\vec{r_2})$ can be said to be roughly uncorrelated when $|\vec{r_1} - \vec{r_2}| > \xi$. If $l \gg \xi$, then $\tilde{h}$ is nearly Gaussian (on account of the central limit theorem), so one can find the imbalance of maxima and minima using the approximation.

The second-order cumulant of $\tilde{h}$ can be expressed as
\begin{equation}
  C_2(\tilde{h}(\vec{r_1}), \tilde{h}(\vec{r_2})) = \frac1{l^4} \iint \rd^2\vec{R_1} \rd^2\vec{R_2} \, C_2(h(\vec{R_1}), h(\vec{R_2})) f\Big(\frac{\vec{R_1}-\vec{r_1}}{l}\Big) f\Big(\frac{\vec{R_2}-\vec{r_2}}{l}\Big).
\end{equation}
Since $C_2(h(\vec{R_1}), h(\vec{R_2}))$ is only appreciable when $|\vec{R_1} - \vec{R_2}| < \xi \ll l$, the approximation $f((\vec{R_2}-\vec{r_2})/l) \approx f((\vec{R_1}-\vec{r_2})/l)$ can be applied. This gives
\begin{equation}
  C_2(\tilde{h}(\vec{r_1}), \tilde{h}(\vec{r_2})) = \frac1{l^4} \iint \rd^2\vec{R_1} \rd^2\vec{a} \, C_2(h(\vec{0}), h(\vec{a})) f\big(\frac{\vec{R_1}-\vec{r_1}}{l}\big) f\big(\frac{\vec{R_1}-\vec{r_2}}{l}\big),
\end{equation}
where $\vec{a} = \vec{R_2}-\vec{R_1}$ and use was made of the homogeneity of $h$. This integration can now be split into two parts:
\begin{align}
  C_2(\tilde{h}(\vec{r_1}), \tilde{h}(\vec{r_2}))	& = \frac1{l^4}	 \int \rd^2\vec{a} \, C_2(h(\vec{0}), h(\vec{a})) \int \rd^2\vec{R_1} f\big(\frac{\vec{R_1}-\vec{r_1}}{l}\big) f\big(\frac{\vec{R_1}-\vec{r_2}}{l}\big) \nl
    							& = \frac1{l^2} I_2 K_2(\frac{\vec{r_1}}{l}, \frac{\vec{r_2}}{l}),
\end{align}
where
\begin{equation}
  I_2 \equiv \int \rd^2\vec{a} \, C_2(h(\vec{0}), h(\vec{a})),
\end{equation}
and
\begin{equation}
  K_2(\vec{\rho_1}, \vec{\rho_2}) \equiv \int \rd^2\vec{v} \, f(\vec{v}-\vec{\rho_1}) f(\vec{v}-\vec{\rho_2}).
\end{equation}
For the third-order correlation an analogous derivation can be made. The result is
\begin{equation}
  C_3(\tilde{h}(\vec{r_1}), \tilde{h}(\vec{r_2}), \tilde{h}(\vec{r_3})) = \frac1{l^4} I_3 K_3(\frac{\vec{r_1}}{l}, \frac{\vec{r_2}}{l}, \frac{\vec{r_3}}{l}),
\end{equation}
with
\begin{equation}
  I_3 \equiv \iint \rd^2\vec{a} \rd^2\vec{b} \, C_3(h(\vec{0}), h(\vec{a}), h(\vec{b})),
\end{equation}
and
\begin{equation}
  K_3(\vec{\rho_1}, \vec{\rho_2}, \vec{\rho_3}) \equiv \int \rd^2\vec{v} \, f(\vec{v}-\vec{\rho_1}) f(\vec{v}-\vec{\rho_2}) f(\vec{v}-\vec{\rho_3}).
\end{equation}
Note in particular that $I_2$ and $I_3$ depend on $h$ only, whereas $K_2$ and $K_3$ depend only on $f$. Also note that none of these terms depends on $l$.

For the correlations, the $z$- and $z^*$-derivatives, as used in Eq.~\eqref{eq:cg_beta}, act only on $K_2(\frac{\vec{r_1}}{l}, \frac{\vec{r_2}}{l})$ and $K_3(\frac{\vec{r_1}}{l}, \frac{\vec{r_2}}{l}, \frac{\vec{r_3}}{l})$. Each derivative introduces a factor $1/l$ as a result of the chain rule. It can therefore already be deduced how the relevant correlations scale with $l$:
\begin{equation}
  \sigma \sim l^{-4},  \qquad  \alpha \sim l^{-6},  \qquad  \beta \sim l^{-8},  \qquad  \gamma, \delta \sim l^{-10},
\end{equation}
and therefore $\dlta n \sim 1/l$. More precisely, we have the following:
\begin{equation}
  \lim_{l/\xi \rightarrow \infty} \dlta n = \frac{c_h c_f}{l},
  \label{eq:maxmin_limit}
\end{equation}
where $c_h = I_3/I_2^{3/2}$ is a parameter that depends on the statistics of $h$ only, and $c_f$ is a parameter that depends on the coarse-graining function $f$ only.





\section{Example}
\label{sec:example}

\subsection{Large scale limit}
\label{subsec:limit}

As an example, let us consider the coarse-grain function from Eq.~\eqref{eq:kpzcg}. This gives
\begin{subequations}
  \begin{align}
    K_2(\vec{\rho_1}, \vec{\rho_2})			& = \frac1{8\pi} e^{ -\tfrac18(\vec{\rho_1}-\vec{\rho_2})^2 }, \\
    K_3(\vec{\rho_1}, \vec{\rho_2}, \vec{\rho_3})	& = \frac1{48\pi^2} e^{ -\tfrac14 \big( \vec{\rho_1}^2+\vec{\rho_2}^2+\vec{\rho_3}^2 - \tfrac13(\vec{\rho_1}+\vec{\rho_2}+\vec{\rho_3})^2 \big) }.
  \end{align}
\end{subequations}
Applying the method as exemplified in Eq.~\eqref{eq:cg_beta}, we obtain
\begin{equation}
  \beta = -\frac{I_3}{6912 \pi l^8}.
\end{equation}
Calculating all relevant cumulants ultimately leads to
\begin{equation}
  \dlta n = \frac{2^{9/2}}{3^{11/2} \pi} \frac{I_3}{I_2^{3/2}} \frac1{l},
\end{equation}
for the KPZ-inspired Gaussian coarse-grain function Eq.~\eqref{eq:kpzcg}.

To get the parameter $c_h$, we use a non-Gaussian field of the form $h = H + \eps H^2$, where $H$ is Gaussian field with a given two-point correlation function $\la H(\vec{r_1}) H(\vec{r_2}) \ra$, and $\eps$ is a small constant. The two-point correlation function of $h$ differs from that of $H$ only in second order of $\eps$ -- this difference will be ignored.

The third-order cumulant of $h$ is zero in leading order, since $H$ is Gaussian, but in first order we find
\begin{equation}
  \begin{split}
    C_3(h(\vec{r_1}), h(\vec{r_2}), h(\vec{r_3})) =\:	& C_3(\eps H(\vec{r_1})^2, H(\vec{r_2}), H(\vec{r_3})) \\
    							& + C_3(H(\vec{r_1}), \eps H(\vec{r_2})^2, H(\vec{r_3})) \\
    							& + C_3(H(\vec{r_1}), H(\vec{r_2}), \eps H(\vec{r_3})^2)
  \end{split}
  \label{eq:3ptcorr}
\end{equation}
This can be expanded with the help of Wick's theorem, e.g.:
\begin{align}
  & C_3(H(\vec{r_1})^2, H(\vec{r_2}), H(\vec{r_3})) \nl
  & \qquad = \la H(\vec{r_1})^2 H(\vec{r_2}) H(\vec{r_3}) \ra - \la H(\vec{r_1})^2 \ra \la H(\vec{r_2}) H(\vec{r_3}) \ra \nl
  & \qquad = 2 \la H(\vec{r_1}) H(\vec{r_2}) \ra \la H(\vec{r_1}) H(\vec{r_3}) \ra,
  \label{eq:wick}
\end{align}
Therefore, the result is
\begin{IEEEeqnarray}{ll}
  \mc{2}{l}{ C_3(h(\vec{r_1}), h(\vec{r_2}), h(\vec{r_3})) } \nl
  \qquad = 2\eps \big(	& \la H(\vec{r_1}) H(\vec{r_2}) \ra \la H(\vec{r_1}) H(\vec{r_3}) \ra + \la H(\vec{r_2}) H(\vec{r_3}) \ra \la H(\vec{r_2}) H(\vec{r_1}) \ra \nl
  			& + \la H(\vec{r_3}) H(\vec{r_1}) \ra \la H(\vec{r_3}) H(\vec{r_2}) \ra \big)
\end{IEEEeqnarray}

If we consider the Gaussian two-point correlation function
\begin{equation}
  \la H(\vec{r_1}) H(\vec{r_2}) \ra = e^{ -\tfrac12 k_0^2 (\vec{r_1}-\vec{r_2})^2 },
\end{equation}
-- which corresponds to the spectrum $A(k) \sim \exp(-k^2/(4k_0^2))$ -- we get
\begin{equation}
  I_2 = \int \rd^2\vec{a} \, C_2(0,\vec{a}) = \int \rd^2\vec{a} \, \la H(0) H(\vec{a}) \ra = \frac{2\pi}{k_0^2},
\end{equation}
and
\begin{align}
  I_3	& = \iint \rd^2\vec{a} \rd^2\vec{b} \, C_3(h(\vec{0}), h(\vec{a}), h(\vec{b})) \nl
  	& = \iint \rd^2\vec{a} \rd^2\vec{b} \, 2\eps \big( e^{-\tfrac12 k_0^2 (\vec{a}^2+\vec{b}^2)} + e^{-\tfrac12 k_0^2 (\vec{a}^2 + (\vec{a}-\vec{b})^2)} + e^{-\tfrac12 k_0^2 (\vec{b}^2 + (\vec{a}-\vec{b})^2)} \big) \nl
  	& = \frac{24 \pi^2 \eps}{k_0^4}.
\end{align}
This gives $c_h = I_3/I_2^{3/2} = 6\sqrt{2\pi}\eps/k_0$.

Combined with the coarse-grain function as given above, we thus get
\begin{equation}
  \dlta n \rightarrow \frac{c_f c_h}{l} = \frac{64}{81\sqrt{3\pi}k_0} \frac{\eps}{l}.
  \label{eq:maxmin_limit_ex}
\end{equation}

\subsection{Analytic result}
\label{subsec:analytic}

The separation of the dependence on $f$ and $h$, as displayed in Eq.~\eqref{eq:maxmin_limit}, is only valid in the limit of $l \gg \xi$. In general however, $f$ and $h$ can no longer be treated separately. Only in very specific cases is it possible to calculate $\dlta n$ for arbitrary $l$. Not coincidentally, the $f$ and $h$ chosen in the previous section allow precisely this.

%

For example, the exact expression for $\beta$ is
\begin{IEEEeqnarray}{rLl}
  \beta	& \mc{2}{L}{ = \la \tilde{h}_z \tilde{h}_{z^*} \tilde{h}_{zz^*} \ra } \nl
  	& = \partial_{z_1} \partial_{z_2^*} \partial_{z_3} \partial_{z_3^*} \iiint	& \rd^2\vec{u_1} \rd^2\vec{u_2} \rd^2\vec{u_3} \, K(\vec{u_1}) K(\vec{u_2}) K(\vec{u_3}) \nl
  	&										& \la h(\vec{r_1}+\vec{u_1}) h(\vec{r_2}+\vec{u_2}) h(\vec{r_3}+\vec{u_3}) \ra \Big|_{\vec{r_1}=\vec{r_2}=\vec{r_3}}.
\end{IEEEeqnarray}
The three-point correlation can be expanded in the same way as before (see Eq.~\eqref{eq:3ptcorr} and Eq.~\eqref{eq:wick}). The final result is
\begin{equation}
  \beta = -\frac{k_0^4 \eps}{2(1+2k_0^2l^2)^2(1+6k_0^2l^2)^2},
\end{equation}

Determining and combining all the correlations gives the following result, which is exact with respect to $l$ but still perturbative with respect to $\eps$:
\begin{equation}
  \dlta n = \frac{ 64 a^3 (1+4a)^{7/2} \eps }{ \sqrt{3\pi} (1+2a)^3 (1+6a)^4 }
  \label{eq:exact}
\end{equation}
where $a \equiv k_0^2 l^2$. In the limit of large $l$ ($a \gg 1$) we find that it matches the perturbative result.

One may also note that Eq.~\eqref{eq:exact} matches the result from the deterministic KPZ equation for the Gaussian spectrum \cite{cite:paper_kpz}, following the substitution $\nu t \rightarrow l^2$.

\section{Numerical tests}
\label{sec:num}

\subsection{Setup}
\label{subsec:setup}

The validity of Eq.~\eqref{eq:exact} was checked using computer simulations. The setup of the simulations and the identification of the extrema is identical to the process outlined in \cite{cite:paper_local}.

For each data point -- corresponding to a particular value of $l$ -- thousands of Gaussian fields $H(\vec{r})$ were generated, following Eq.~\eqref{eq:gaussian}. In each case, hundreds of waves were summed, each with an amplitude in accordance with the desired spectrum and a random value for the phase. The values of $H$ were evaluated for the points of a square grid of size $L$. Periodic boundary conditions were enforced to reduce finite size effects, which entail that the components of the selected wave vectors $k$ were all multiples of $\frac{2\pi}{L}$.

To this field, the perturbation $\eps H^2$ was added, with $\eps = 0.1$. This new field $h$ was then coarse-grained, after which the extrema were identified. For this identification, the first and second derivatives of the coarse-grained field $\tilde{h}$ were used, which were determined by calculating the derivatives of the original Gaussian field $H$ and letting them undergo the equivalent process.

\subsection{Results}
\label{subsec:results}

\begin{figure}
  \centering
  \includegraphics[width=.7\textwidth]{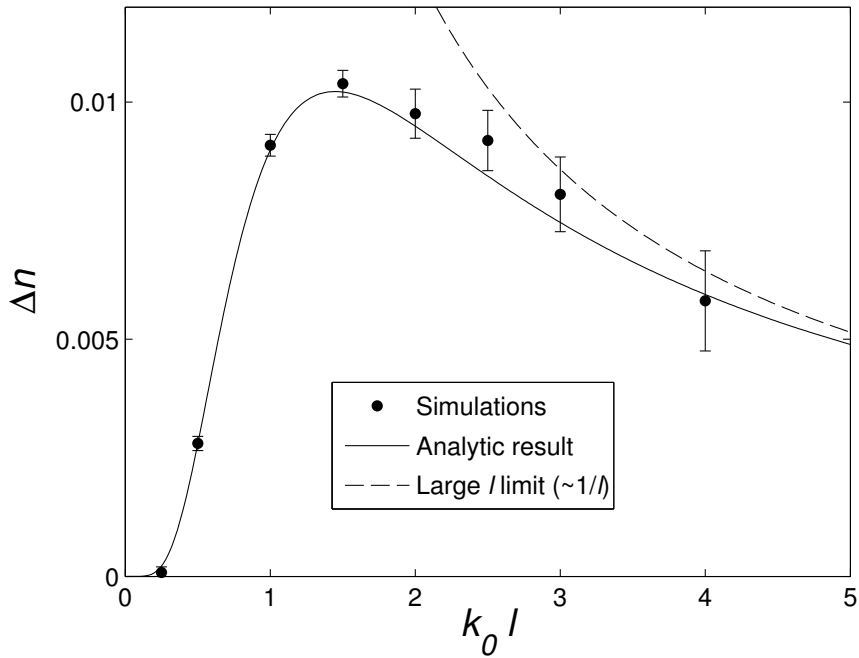}
  \caption{The imbalance between maxima and minima $\dlta n$ for a field $h = H + \eps H^2$, where $H$ is a Gaussian field with a Gaussian power spectrum (with typical wavelength $k_0^{-1}$), coarse-grained with a Gaussian function $f(r) = \exp(r^2/(4l^2))$. The solid line is the exact result Eq.~\eqref{eq:exact} (perturbative in $\eps$ but not in $l$), while the data points stem from simulations. The dashed line is the theoretical result for large $l$, Eq.~\eqref{eq:maxmin_limit_ex}.}
  \label{fig:maxmin}
\end{figure}

Figure~\ref{fig:maxmin} shows the theoretical result, as well as results from simulations at various values of $l$. As can be seen, there is an excellent agreement between the two. Also shown is the prediction of Eq.~\eqref{eq:maxmin_limit_ex}, illustrating the large $l$ limit, which matches well for $k_0 l \gg 1$.

An interesting point is that, for no coarse-graining at all, the imbalance is very close to zero. This general feature of local perturbations of the type $h = H + f(H)$ was already established in detail in \cite{cite:paper_local}. Measuring the imbalance between maxima and minima thus does not reveal the non-Gaussianity of $h$. However, it is clear from figure~\ref{fig:maxmin} that coarse-graining may significantly increase the imbalance to measurable values, thereby not only granting the possibility of detecting non-Gaussianity, but also potentially identifying the size and type of the perturbation.

\subsection{Large scale coarse-graining}
\label{subsec:largescale}

It is difficult to test the formula for a large coarse-graining (Eq.~\eqref{eq:maxmin_limit}) accurately in a numerical setting, since the size of the system is limited.

As said, the periodic boundary conditions are enforced by only using waves with wave vectors $\vec{k}$ for which the components are multiples of $\frac{2\pi}{L}$, where $L$ is the system size. As long as $L$ is large, this quantization has a high enough resolution to be of no significant source of error.

Now consider what happens when it is coarse-grained. We already saw in Eq.~\eqref{eq:gsncg} that, effectively, its amplitude spectrum changes:
\begin{equation}
  \tilde{A}(k) = A(k) \Big( \int \rd^2\vec{u} \, K(\vec{u}) e^{i \vec{k} \cdot \vec{u}} \Big) = A(k) \Big( \int \rd^2\vec{\rho} \, f(\rho) e^{i l \vec{k} \cdot \vec{\rho}} \Big).
\end{equation}
In the limit that $lk \gg 1$, the phase factor causes the integral to vanish. Hence, for large $l$, only the waves with small wave vector $k$ (in the order of $1/l$ or less) prevail. This is the technical justification of the statement that coarse-graining causes a field to become smoother, and thus dominated by long waves.

However, in combination with the periodic boundary conditions, this means that -- in the case that $l$ becomes comparable to $L$ -- there are only a few wave vectors left that are of importance from the coarse-graining point of view. The accuracy with which the simulated coarse-grained field represents an actual field thus becomes compromised. Therefore, $L$ should be larger than $l$. Increasing $L$ however naturally increases computation time. As a result, probing large coarse-grain scales indirectly requires a lot of computation time, making it difficult to properly explore the regime in which the imbalance $\dlta n$ decays as $1/l$.

\section{Conclusions}
\label{sec:concl}

Coarse-graining a non-Gaussian field has the effect of giving it Gaussian characteristics, as the coarse-graining scale goes to infinity. More precisely, when this scale $l$ is significantly larger than the correlation length of the field, the imbalance between maxima and minima -- which is zero for Gaussian fields -- scales as $1/l$. The corresponding constant factor can be written as the product of two independent scalars: one depends on the field only, whereas the other depends on the coarse-graining function only.

Coarse-graining a signal on purpose can also be useful, because the imbalance between maxima and minima depends on the length scale of the coarse-graining for a non-Gaussian field. For example, locally perturbed fields, such as $h = H + \eps H^2$, where $H$ is Gaussian, do not show a significant imbalance between maxima and minima (if the resolution is perfect). However, coarse-graining -- which would not produce an effect for Gaussian fields -- creates an imbalance allowing $\eps$ to be measured. In general, coarse-graining by various amounts can give a multitude of data that can be used to shed light on some unknown parameters of the perturbation.

\bibliography{nongauss}

\end{document}